\documentclass{article}
\usepackage{amsmath}
\usepackage[dvips]{graphics}
\usepackage{latexsym}
\begin{document}
\newcommand{\bea}{\begin{eqnarray*}}
\newcommand{\eea}{\end{eqnarray*}}
\newcommand{\bean}{\begin{eqnarray}}
\newcommand{\eean}{\end{eqnarray}}
\newcommand{\eqs}[1]{Eqs. (\ref{#1})}
\newcommand{\eq}[1]{Eq. (\ref{#1})}
\newcommand{\meq}[1]{(\ref{#1})}
\newcommand{\fig}[1]{Fig. \ref{#1}}

\newcommand{\tri}{\delta}
\newcommand{\grad}{\nabla}
\newcommand{\pa}{\partial}
\newcommand{\pf}[2]{\frac{\pa #1}{\pa #2}}
\newcommand{\cla}{{\cal A}}
\newcommand{\aqt}{\frac{1}{4}\theta}

\newcommand{\oh}{\frac{1}{2}}
\newcommand{\hsp}{\hspace{0.1mm}}
\newcommand{\spa}{\hspace{3mm}}
\newcommand{\hst}{\ \ \ }
\newcommand{\upd}[2]{^#1\hsp_#2}
\newcommand{\eqn}{&=&}
\newcommand{\non}{\nonumber \\}
\newcommand{\ppa}[2]{\left(\frac{\partial}{\partial #1}\right)^{#2}}
\newcommand{\pp}[2]{\frac{\partial #1}{\partial #2}}
\newcommand{\rn}{Reissner-Nordstr\"om}
\newcommand{\vphi}{\varphi}
\newcommand{\kss}{\left[K_\theta^\theta \right]}
\newcommand{\ktt}{\left[K_\tau^\tau \right]}
\newcommand{\heq}{{\hat =}}
\newcommand{\dE}{\delta E}
\newcommand{\ppb}[3]{\frac{\partial #1}{\partial #2}\Big|_{#3}}

\title{A general maximum entropy principle for self-gravitating perfect fluid}
\author{ Sijie Gao\footnote{ Email: sijie@bnu.edu.cn} \\
Department of Physics, Beijing Normal University,\\
Beijing 100875, China}
\maketitle

\begin{abstract}
We consider a self-gravitating system consisting of perfect fluid with spherical symmetry. Using the general expression of entropy density, we extremize the total entropy $S$ under the constraint that the total number of particles is fixed. We show that extrema of $S$ coincides precisely with the relativistic Tolman-Oppenheimer-Volkoff equation of hydrostatic equilibrium. Furthermore, we apply the maximum entropy principle to a charged perfect fluid and derive the generalized Tolman-Oppenheimer-Volkoff equation. Our work provides a strong evidence for the fundamental relationship between general relativity and ordinary thermodynamics.\\

\end{abstract}

\section{Introduction}
In the past few decades, research in general relativity has suggested a very deep connection between gravitation and thermodynamics. The four laws of black hole mechanics were originally derived from the Einstein equation at the purely classical level\cite{bardeen1973}. The discovery of the Hawking radiation \cite{hawking} allows a consistent interpretation of the laws of black hole mechanics as the ordinary laws of thermodynamics. By turning the logic around, Jacobson \cite{ted} showed that the Einstein equation may be derived from the first law of local Rindler horizons.   Inspired by Jacobson's work, a lot of efforts have been made to derive the dynamical equations from black hole thermodynamics \cite{gong}-\cite{c4}.  In fact, this idea can be traced back even before the establishment of black hole mechanics. In 1965, Cocke \cite{cocke} proposed a maximum entropy principle for self-gravitating fluid spheres. Let $S$ be the  total entropy of spherically symmetric perfect fluid.  Cocke showed that the requirement that $S$ be an extremum yields the equation of hydrostatic equilibrium which was originally derived from the Einstein equation. However, a critical assumption in Cocke's derivation is that the fluid is in adiabatic motion so that the total entropy is invariant. By imposing the adiabatic condition, the entropy density $s$ is expressed as the function of the energy density $\rho$, while for a general fluid, $s$ is a function of at least two thermodynamic variables. In my opinion, the variation of $S$ is performed on a spacelike hypersurface and the dynamic revolution of the fluid is irrelevant. Furthermore, variation of $S$ is not consistent with the adiabatic condition. If the entropy is required to be invariant, as indicated by the adiabatic condition, the variation of entropy would be meaningless. Thus, it is not appropriate and consistent to impose the adiabatic condition. In relation to Cocke's work, Sorkin, Wald and Zhang (SWZ)\cite{wald} develop a different entropy principle for radiation. The major difference is that the adiabatic condition was not needed in SWZ's derivation. Moreover only the Einstein constraint equation was used in the proof while Cocke used both the constraint equation and the radial-radial component of Einstein's equation. However, SWZ's discussion was restricted to radiation for which the thermodynamic relations can be expressed explicitly and the entropy density only depends on one thermodynamic variable. It is important to know whether SWZ's treatment can be generalized to an arbitrary perfect fluid. In this paper, we prove a maximum entropy principle for a general self-gravitating perfect fluid. The new arguments used in our proof are as follows. First, we use the Gibbs-Duhem relation as the expression of entropy density for a general fluid. Second, our maximum entropy principle is under the constraint that the total number of particles is invariant. Consequently, the method of Lagrange multipliers plays an important role in our derivation. Third, in addition to the Einstein constraint equation, we only make use of the ordinary thermodynamic relations to derive the Tolman-Oppenheimer-Volkoff (TOV) equation. No other assumptions are needed. Finally, we extend our treatment to a general charged fluid. With modified arguments, we derive the generalized TOV equation for a charged fluid.

\section{Review of SWZ's derivation on self-gravitating radiation}

Since our work is closely related to SWZ's prescription, we shall give a brief review on the derivation in \cite{wald}. Consider a spherical box of radiation having total energy $M$ and confined within a radius $R$. For thermal radiation, the pressure $p$ and energy density $\rho$ satisfy the equation of state
\bean
p=\frac{1}{3}\rho \,,
\eean
and then the stress energy tensor is given by
\bean
T_{ab}=\rho u_au_b+\frac{1}{3}\rho(g_{ab}+u_au_b)\,,
\eean
where $u^a$ is the 4-velocity of the local rest frame of the radiation.  In terms of the locally measured temperature $T$, the energy density $\rho$ and entropy density $s$ are given by
\bean
\rho\eqn b T^4 \,,\\
s\eqn\frac{4}{3}bT^3 \,,
\eean
where $b$ is a constant. So $s$ can also be expressed as
\bean
s=\alpha\rho^{3/4}
\eean
with $\alpha=\frac{4}{3}b^{1/4}$.
As shown by SWZ, the extrema of the total entropy $S$ corresponds to a static spacetime metric
\bean
ds^2=g_{tt}(r)dt^2+\left[1-\frac{2m(r)}{r}\right]^{-1} dr^2+r^2 d\Omega^2\,.
\eean
The constraint equation, which is obtained from the time-time component of the Einstein equation, yields
%************rhm
\bean
\rho=\frac{m'(r)}{4\pi r^2} \label{rhm}\,.
\eean
Thus, $m(r)$ is a mass function.

Let the gas be confined in the region $r\leq R$. Then the total entropy is given by
\bean
S\eqn 4\pi \int_0^R s(r)\left[1-\frac{2m(r)}{r}\right]^{-1/2}r^2 dr\non
\eqn 4\pi\alpha\int_0^R\rho^{3/4}\left[1-\frac{2m(r)}{r}\right]^{-1/2}r^2 dr\non
\eqn (4\pi)^{1/4}\alpha \int_0^R \left[\frac{1}{r^2}m'(r)\right]^{3/4}\left[1-\frac{2m(r)}{r}\right]^{-1/2}r^2 dr\,.
\eean
Our task is to find a function $m(r)$ such that the total entropy $S$ is extremized. Note that
the total mass $M$ within $R$ is
%*************bign
\bean
M=4\pi\int_0^R\rho(r)r^2 dr=m(R)\label{bign}\,,
\eean
and obviously
\bean
m(0)=0\,.
\eean
Hence, all the variations must satisfy
%*************conm
\bean
\delta m(0)=\delta m(R)=0 \,.\label{conm}
\eean
By using this condition, the extrema of $S$ is equivalent to the Euler-Lagrange equation
%********************emp
\bean
\frac{d}{dr}\left(\pp{L}{m'}\right)-\pp{L}{m}= 0  \label{emp}
\eean
for the Lagrangian
\bean
L=(m')^{3/4}\left[1-\frac{2m(r)}{r}\right]^{-1/2}r^{1/2}\,.
\eean
By straightforward calculation, \eq{emp} yields
%**********************mppr
\bean
-\frac{3}{16}m'' r^2+\frac{3}{8}m''mr+\frac{3}{8}m'r-\frac{1}{4}m'^2r-\frac{3}{2}m'm=0 \,. \label{mppr}
\eean
By substituting \eq{rhm}, one can show that \eq{mppr} is equivalent to
%*******************tov
\bean
\frac{d}{dr}\left(\rho/3\right)=-\frac{(\rho+\rho/3)[m(r)+4\pi r^3 (\rho/3)]}{r[r-2m(r)]}\,. \label{tov}
\eean
Since $p=\rho/3$ for radiation, we see immediately that \eq{tov} is just the relativistic Tolman-Oppenheimer-Volkoff equation.

\section{Maximum entropy principle for perfect fluid}
To generalize SWZ's prescription to an arbitrary perfect fluid, we first need to find a formula for entropy density $s$. Because radiation has a vanishing chemical potential, its entropy density depends on only one thermodynamic variable, e.g. $T$ or $\rho$. For fluids consisting of particles, there are at least two independent variables. We start with the familiar first law
%***************fign
\bean
dS=\frac{1}{T}dE+\frac{p}{T}dV-\frac{\mu}{T}dN \,,\label{fign}
\eean
where $S,E,N$ represent the total entropy, energy and particle number within the volume $V$.
Write \eq{fign} in terms of densities
\bean
d(sV)=\frac{1}{T}d(\rho V)+\frac{p}{T}dV-\frac{\mu}{T}d(nV)\,.
\eean
By expansion, we have
%************sdv
\bean
sdV+Vds=\frac{1}{T}\rho dV+V d\rho+\frac{p}{T}dV-\frac{\mu}{T}ndV-\frac{\mu}{T}Vdn \,.\label{sdv}
\eean
Applying \eq{fign} to a unit volume, we find
%*************flaw
\bean
ds=\frac{1}{T}d\rho-\frac{\mu}{T}dn \,.\label{flaw}
\eean
Combining \eqs{sdv} and \meq{flaw}, we arrive at the integrated form of the  Gibbs-Duhem relation \cite{me}
%**************str
\bean
s=\frac{1}{T}(\rho+p-\mu n)\,. \label{str}
\eean
 To derive this formula, we only used the first law of the ordinary thermodynamics. So it is a general expression for perfect fluid.
We treat $(\rho, n)$ as two independent variables, e.g.,
\bean
s=s(\rho,n), \ \mu=\mu(\rho,n),\ \ p=p(\rho,n)\,.
\eean

For example, the thermodynamic quantities for a monatomic ideal gas are given by \cite{landau}
\bean
\rho\eqn\frac{3}{2}nkT \,,\\
p\eqn nkT \,,\\
s\eqn\frac{3}{2}nk\ln T-nk\ln n+\frac{3}{2}nk\left[\frac{5}{3}+\ln\left(\frac{2\pi mk}{h^2}\right)\right]\,.
\eean

Our task is to extremize the total entropy
\bean
S=4\pi\int_0^R s(r)\left[1-\frac{2m(r)}{r}\right]^{-1/2}r^2 dr\,.
\eean
In addition to the constraint \eq{conm}, it is natural to require the total number of particles
%*************tn
\bean
N=4\pi\int_0^R n(r)\left[1-\frac{2m(r)}{r}\right]^{-1/2}r^2 dr \label{tn}
\eean
to be invariant, i.e.,
\bean
\delta N=0\,.
\eean

Following the standard method of Lagrange multipliers, the equation of variation becomes
\bean
\delta S+\lambda \delta N=0\,.
\eean
Define the ``total Lagrangian'' by
%*************bigl
\bean
L(m,m',n)=s(\rho(m'),n)\left[1-\frac{2m(r)}{r}\right]^{-1/2}r^2+\lambda n(r)\left[1-\frac{2m(r)}{r}\right]^{-1/2}r^2 \,.\label{bigl}
\eean
Now the constrained Euler-Lagrange equation is given by
%*********************bln blm
\bean
\pp{L}{n}\eqn 0 \,, \label{bln} \\
\frac{d}{dr}\pp{L}{m'}+\pp{L}{m}\eqn 0 \,.\label{blm}
\eean
Thus, \eq{bln} yields
%*************plpn
\bean
\pp{s}{n}+\lambda=0 \,.
\eean
Using \eq{flaw}, we have
%*******************mtl
\bean
-\frac{\mu}{T}+\lambda=0 \,,\label{mtl}
\eean
which shows that $\frac{\mu}{T}$ must be a constant for self-gravitating fluid.

From \eq{bigl}, we have
%*************lma
\bean
\pp{L}{m}=r\left(1-\frac{2m}{r}\right)^{-3/2}(n\lambda+s) \label{lma}\,,
\eean
and
\bean
\pp{L}{m'}=\pp{s}{m'}r^2\left(1-\frac{2m}{r}\right)^{-1/2}\,.
\eean
Here
\bean
\pp{s}{m'}=\pp{s}{\rho}\pp{\rho}{m'}=\frac{1}{T}\frac{1}{4\pi r^2}\,,
\eean
where \eqs{rhm} and \meq{flaw} have been used.
Hence
\bean
\pp{L}{m'}=\frac{1}{4\pi T}\left(1-\frac{2m}{r}\right)^{-1/2}\,,
\eean
and
\bean
\frac{d}{dr}\pp{L}{m'}=\frac{T(m'r-m)-r(r-2m)T'}{4\pi T^2(r-2m)^{3/2}r^2}\,.
\eean
Using \eqs{mtl} and \meq{str}, \eq{lma} becomes
%*************lmb
\bean
\pp{L}{m}=r\left(1-\frac{2m}{r}\right)^{-3/2}\left(\frac{\rho+p}{T}\right) \label{lmb}\,.
\eean
So the Eular-Lagrange \eq{blm} yields
\bean
(4\pi p r^3+m)T+(r-2m)rT'=0 \label{ell}\,.
\eean

The constraint \eq{mtl} yields
%*************mmtp
\bean
\mu'=\lambda T'\,. \label{mmtp}
\eean

Rewrite \eq{str} as
\bean
p=Ts+\mu n-\rho\,.
\eean
The differential of $p$ is
\bean
dp=Tds+sdT+\mu dn+nd\mu-d\rho\,.
\eean

By substituting \eq{flaw}, we have
%***********pst
\bean
dp=sdT+nd\mu \label{pst}\,.
\eean
It follows immediately  that
%**********pps
\bean
p'(r)=sT'(r)+n\mu'(r) \,.\label{pps}
\eean

Substituting \eqs{mtl}, \meq{str} and \meq{mmtp} into \eq{pps}, we have
%************tpf
\bean
T'=\frac{T}{p+\rho}p'(r) \label{tpf}\,.
\eean
Substituting \eq{tpf} into \eq{ell}, we obtain the desired TOV equation
\bean
p'=-\frac{(p+\rho)(4\pi r^3 p+m)}{r(r-2m)}\,.
\eean

\section{Maximum entropy principle for charged fluid}
For a charged fluid, the local thermodynamic relations remain unchanged. For example, we can still use \eq{str} as the expression of entropy density. But the presence of charge will change the distribution of the fluid in spacetime.

In coordinates $(t,r,\theta,\phi)$, assume that a spherically
symmetric charged fluid has the line element
%*********************metric
\bean
ds^2=g_{tt}(r)dt^2+\left[1-\frac{2m(r)}{r}+\frac{Q^2(r)}{r^2}\right]^{-1}dr^2+r^2 d\theta^2+r^2\sin^2\theta
d\phi^2\,.
\label{metric}
\eean
Here $Q(r)$ is defined as the total charge up to the radius $r$ and $m(r)$ will be determined later. The matter field consists of a charged  fluid. Let
%************************ua
\bean
u^{a}=\frac{1}{\sqrt{-g_{tt}}}\ppa{t}{a}\label{ua}
\eean
be the four-velocity of the fluid. Then the
total stress-energy tensor can be written as
%*************************tt
\bean
T_{ab}=\tilde T_{ab}+T^{EM}_{ab} \label{tt}\,,
\eean
where
%*************tp
\bean
\tilde T_{ab}=\rho u^au^b+p(g_{ab}+u_au_b) \label{tp}\,,
\eean
%**************tem
\bean
T^{EM}_{ab}=\frac{1}{4\pi}\left(F_a^{\
c}F_{bc}-\frac{1}{4}g_{ab}F^{cd}F_{cd} \right) \label{tem}\,.
\eean

The electromagnetic field $F_{ab}$ satisfies the Maxwell's equations
%****************dF
\bean
\grad_b F^{ab}=4\pi j^a=4\pi \rho_e u^a \label{dF}\,,
\eean
and
%**********************asf
\bean
\grad_{[a}F_{bc]}=0 \label{asf}\,,
\eean
where $\rho_e$ is the charge density measured by the comoving observers.  By using the identity $\Gamma^a_{a\mu}=\frac{\partial}{\partial
x^{\mu}}\ln \sqrt{-g}$ (see \cite{waldbook})
\eq{dF} becomes
%****************ndf
\bean
\frac{\partial}{\partial
x^{\nu}}\left[\sqrt{-g}F^{\mu\nu}\right]=4\pi j^{\mu}\sqrt{-g}
\label{ndf}\,.
\eean
Because of spherical symmetry, the only nonvanishing components of
$F^{ab}$ are $F^{tr}(r)=-F^{rt}(r)$. Thus, \eq{ndf} yields
%*************fp
\bean
\partial_r (r^2 \sqrt{-g_{tt}g_{rr}}F^{tr})=4\pi j^t r^2
\sqrt{-g_{tt}g_{rr}}=4\pi \rho_e r^2
\sqrt{g_{rr}} \label{fp}\,,
\eean
where \eqs{ua} and \meq{dF} have been used in the last step. So by definition, the function $Q(r)$ in \eq{metric} can be written as
%****************cq
\bean
Q(r)=\int_0^r 4\pi r'^2\sqrt{g_{rr}}\rho_e dr' \label{cq}\,.
\eean
By comparing \eqs{cq} and \meq{fp}, one finds immediately that
%****************ftr
\bean
F^{tr}=\frac{1}{r^2\sqrt{-g_{tt}g_{rr}}}Q(r) \label{ftr}
\eean
is a solution of \eq{fp}.

Then the time-time component of Einstein's equation gives
%******************mptt
\bean
m'(r)=4\pi r^2 \rho+\frac{QQ'}{r} \label{mptt}\,.
\eean
This formula is consistent with the result in \cite{bekenstein}. Now we derive the hydroelectrostatic equation from the maximum entropy principle. The total entropy of matter takes the form
\bean
S=\int_0^R s(r)\left[1-\frac{2m}{r}+\frac{Q^2}{r^2}\right]^{-1/2}r^2dr\,.
\eean
For simplicity, we assume all the particles have the same charge $q$. Thus, the charge density is proportional to the  particle number density $n$
%***********reqn
\bean
\rho_e=q n \label{reqn}\,.
\eean
Together with \eq{cq}, we have
%*********************rhoe
\bean
n=\frac{Q'}{4\pi r^2 q}\left[1-\frac{2m}{r}+\frac{Q^2}{r^2}\right]^{1/2}\label{rhoe}\,.
\eean
Now we treat $Q(r),Q'(r)$ as independent variables in the Lagrangian formalism. So the Lagrangian is written as
%*********************lmm
\bean
L(m,m',Q,Q')=  s\left[1-\frac{2m}{r}+\frac{Q^2}{r^2}\right]^{-1/2}r^2 \label{lmm}\,.
\eean
The conservation of particle number $N$ is equivalent to the conservation of charge with the radius $R$. Now the constraints are
\bean
m(0)=Q(0)=0,\ \ m(R)=constant, \ \ Q(R)=constant\,.
\eean
With these constraints, the extrema of $S$ leads to the following Euler-Lagrange equations
%************qel mel
\bean
\frac{d}{dr}\pp{L}{Q'}+\pp{L}{Q}\eqn 0 \label{qel} \\
\frac{d}{dr}\pp{L}{m'}+\pp{L}{m}\eqn 0 \label{mel}
\eean

To calculate the Euler-Lagrange equations, we first note that
%**************srn
\bean
s=s(\rho, n)=s(\rho(m',Q,Q'),n(Q,m,Q')) \label{srn}\,.
\eean
With the help of \eqs{mptt} and \meq{rhoe}, we have
\bean
\pp{s}{Q'}\eqn\pp{s}{\rho}\pp{\rho}{Q'}+\pp{s}{n}\pp{n}{Q'}\\
\eqn -\frac {1}{T}\frac{Q}{4\pi r^3}-\frac{\mu}{T} \frac{1}{q}\frac{1}{4\pi r^2}\left[1-\frac{2m}{r}+\frac{Q^2}{r^2}\right]^{1/2}\,.
\eean

Thus,
%***********lqpa
\bean
\pp{L}{Q'}= -\frac {1}{T}\frac{Q}{4\pi r}\left[1-\frac{2m}{r}+\frac{Q^2}{r^2}\right]^{-1/2}-\frac{\mu}{T} \frac{1}{q}\frac{1}{4\pi}\,. \label{lqpa}
\eean

To calculate $\pp{L}{Q}$, first note that
\bean
\pp{s}{Q}\eqn\pp{s}{\rho}\pp{\rho}{Q}+\pp{s}{n}\pp{n}{Q} \non
\eqn -\frac{1}{T}\frac{Q'}{4\pi r^3}-\frac{\mu}{T}\frac{QQ'}{4\pi q r^4}\left[1-\frac{2m}{r}+\frac{Q^2}{r^2}\right]^{-1/2}\,.
\eean
Then
%******************lqqb
\bean
\pp{L}{Q}=-\frac{4\pi r^2 q QsT+(f qr+\sqrt f Q\mu)Q'}{4\pi r^2 q T f^{3/2}}\,, \label{lqqb}
\eean
where
\bean
f=1-\frac{2m}{r}+\frac{Q^2}{r^2}\,.
\eean

By substituting \eqs{lqpa} and \meq{lqqb}, \eq{qel} becomes
%************qcc
\bean
0\eqn qQ^3 T'+Q[-mqT+qrT-qrTm'+4\pi qr^3sT^2+ \sqrt f rT\mu Q'-2mqrT'\non
&+&qr^2T']+\sqrt f r^2(r-2m)(\mu T'-T\mu')+Q^2[qTQ'+\sqrt f r(\mu T'-T\mu')]\,.\non
&&\label{qcc}
\eean

Using \eq{pps} to eliminate $\mu'$ in \eq{qcc},  we have
%**********tplo
\bean
0\eqn qQ^3T'+\frac{\sqrt f r^2(r-2m)(sTT'+n\mu T'-Tp')}{n}+\frac{Q^2\sqrt f r(sTT'+n\mu T'-Tp')}{n}  \non
&+&qTQ^2Q'+Q[-mqT+qrT-qrTm'+4\pi qr^3sT^2+ \sqrt f rT\mu Q'-2mqrT'+qr^2T']\,.\non
&&\label{tplo}
\eean
Eliminating $s$, $\mu$ and $n$ via \eqs{str} and \meq{rhoe},  we rewrite \eq{tplo} as
%***************tpfl
\bean
0\eqn 4\pi r^3(r^2-2mr+Q^2)(p+\rho)T'-4\pi r^3(r^2-2mr+Q^2)T p'+TQ^2Q'^2\non
&+& QQ' (rT+4\pi r^3 (p+\rho)T+Q^2 T'+r^2T'-mT-2r m T'-rm'T) \label{tpfl}\,.
\eean

Now we begin to calculate  \eq{mel}. Note that
\bean
\pp{s}{m'}=\pp{s}{\rho}\pp{\rho}{m'}=\frac{1}{4\pi r^2T}\,.
\eean
Then
\bean
\pp{L}{m'}=\frac{1}{4\pi r^2T}\left[1-\frac{2m}{r}+\frac{Q^2}{r^2}\right]^{-1/2}r^2\,.
\eean

\eq{srn} yields
\bean
\pp{s}{m}=\pp{s}{n}\pp{n}{m}=\frac{\mu}{T}\frac{Q'}{4\pi r^3 q\sqrt{1-\frac{2m}{r}+\frac{Q^2}{r^2}}}\,.
\eean
Here we have used \eqs{flaw} and \eq{rhoe}.
From \eq{lmm}, we find
\bean
\pp{L}{m}\eqn \pp{s}{m}\left[1-\frac{2m}{r}+\frac{Q^2}{r^2}\right]^{-1/2}r^2+sr \left[1-\frac{2m}{r}+\frac{Q^2}{r^2}\right]^{-3/2} \non
\eqn \frac{\mu}{T}\frac{Q'}{4\pi r^2 q}  \left[1-\frac{2m}{r}+\frac{Q^2}{r^2}\right]^{-1}r+sr \left[1-\frac{2m}{r}+\frac{Q^2}{r^2}\right]^{-3/2}\non
\eqn r\left[1-\frac{2m}{r}+\frac{Q^2}{r^2}\right]^{-3/2}\left[ \frac{\mu}{T}\frac{Q'}{4\pi r^2 q}\left[1-\frac{2m}{r}+\frac{Q^2}{r^2}\right]^{1/2}+s\right]\non
\eqn r\left[1-\frac{2m}{r}+\frac{Q^2}{r^2}\right]^{-3/2}\left[ \frac{\mu n}{T}+s\right]\non
\eqn r\left[1-\frac{2m}{r}+\frac{Q^2}{r^2}\right]^{-3/2} \frac{\rho+p}{T}\,.
\eean
Thus, \eq{mel} becomes
%*******************mfi
\bean
&&Q^2 T-4\pi r^4 T(p+\rho)+m'Tr^2-TrQQ'-rT'Q^2-r^3T'\non
&&-mrT+2mr^2T'=0 \,.\label{mfi}
\eean
Combining \eq{tpfl} and \eq{mfi}, one can eliminate $T'$. Then by substituting \eq{mptt} for $m'$, we finally find
%**************ovg
\bean
p'=\frac{QQ'}{4\pi r^4}-(\rho+p)\left(4\pi rp+\frac{m}{r^2}-\frac{Q^2}{r^3}\right)\left(1-\frac{2m}{r}
+\frac{Q^2}{r^2}\right)^{-1} \,.\label{ovg}
\eean
This is exactly the generalized Oppenheimer-Volkoff equation for charged fluid \cite{bekenstein}.

\section{Conclusions}
By applying the maximum entropy principle to a general self-gravitating fluid, we have derived the TOV equation of hydrostatic equilibrium. We only used the Einstein's constraint equation and ordinary thermodynamic relations. By similar assumptions but more complicated arguments, we have shown that the generalized TOV equation for a charged fluid can also be derived by extremizing the total entropy. The TOV equation is an important equation for self-gravitating system which was originally derived from the Einstein equation. Our results show that the Einstein equation can be derived from ordinary thermodynamic laws. This is direct evidence for the fundamental relationship between gravitation and thermodynamics.

\section*{Acknowledgements}
This research was supported by NSFC Grants No. 10605006, 10975016 and by``The Fundamental Research Funds for the Central Universities."

\end{document}